\documentclass[10pt,twocolumn,3p]{elsarticle}

\usepackage[utf8]{inputenc}
\usepackage[]{hyperref}
\usepackage[]{graphicx}
\usepackage{xcolor}

\usepackage{amsmath,amssymb}

\usepackage{miller}

\newcommand{\etal}{\textit{et al}.}

\newcommand{\degree}{$^{\circ}$}

\journal{Acta Materialia}

\begin{document}

\begin{frontmatter}

\title{Impact of magnetism on screw dislocations in body-centered cubic chromium}

\author[SRMP]{Baptiste Bienvenu}                                                                                                                                                                                                                                                                                            \author[SRMP]{Chu Chun Fu}                                                                                                                                                                                                                                                   
\author[SRMP]{Emmanuel Clouet\corref{CA}}                                                                                                                                                                                                                                      
\cortext[CA]{Corresponding author}                                                                                                                                                                                                                                             
\ead{emmanuel.clouet@cea.fr}                                                                                                                                                                                                                                                   
                                                                                                                                                                                                                                                                               
\address[SRMP]{Université Paris-Saclay, CEA, Service de Recherches de Métallurgie Physique, F-91191 Gif-sur-Yvette, France}                       

\begin{abstract}
The influence of magnetism on the properties of screw dislocations in body-centered cubic chromium is investigated by means of \textit{ab initio} calculations. Screw dislocations having Burgers vectors $1/2\,\hkl<111>$ and \hkl<100> are considered, following experimental observations showing activity for both slip systems. At low temperature, chromium has a magnetic order close to antiferromagnetism along \hkl<100> directions, for which $1/2\,\hkl<111>$ is not a periodicity vector. Hence, dislocations with Burgers vectors $1/2\,\hkl<111>$ generate magnetic faults when shearing the crystal, which constrain them to coexist and move pairwise, leading to dissociated \hkl<111> super-dislocations. On the other side, \hkl<100> is a periodicity vector of the magnetic order of chromium, and no such magnetic fault are generated when \hkl<100> dislocations glide. Dislocation properties are computed in the magnetically ordered and non magnetic phases of chromium for comparison purposes. We report a marginal impact of magnetism on the structural properties and energies of dislocations for both slip systems. The Peierls energy barrier opposing dislocation glide in \hkl{110} planes is comparable for both $1/2\,\hkl<111>$\,\hkl{110} and \hkl<100>\,\hkl{110} slip systems, with lower Peierls stresses in the magnetically ordered phase of chromium.
\end{abstract}


\end{frontmatter}

\section{Introduction}
\label{section:intro}
As a body-centered cubic (BCC) metal, the plastic behavior of chromium (Cr) at low temperature is \textit{a priori} governed by screw dislocations gliding in \hkl{110} planes \cite{anderson2017}. The motion of these dislocations is difficult and needs thermal activation, leading to brittleness of Cr at low temperature \cite{marcinkowski1962,sameljuk1996,gu2004,fritz2016,fritz2017,choi2017,kiener2019}. These dislocations have a Burgers vector corresponding to the smallest periodicity vector of the crystal lattice, $1/2\,\hkl<111>$. However, Cr has a spin-density wave magnetic ground-state, which shows a locally antiferromagnetic ordering along a \hkl<100> direction with a modulation of spin magnitudes \cite{fawcett1988,jacques2009}, and this Burgers vector does not correspond to a periodicity vector of the magnetic order. Hence, when the crystal is sheared by such dislocations, magnetic faults should be generated in the dislocation glide planes, possibly impeding the motion of these line-defects.

Using transmission electron microscopy (TEM), $1/2\,\hkl<111>$ dislocations were found in strained Cr polycrystals \cite{mclaren1964,reid1966,hale1969} at temperatures where magnetic order prevails, below the Néel temperature {of 311\,K}. Slip traces analysis confirm that these dislocations are gliding in \hkl{110} planes \cite{mclaren1964,reid1966,reid1967}, with glide in the \hkl{112} and \hkl{123} planes also observed at high temperature \cite{mclaren1964}. Regarding the disruption of the magnetic order, both Ravlic {\etal} \cite{ravlic2003} and Kleiber {\etal} \cite{kleiber2000} show the existence of AF domains at \hkl{100} surfaces separated by walls using spin-polarized scanning tunneling microscopy at room temperature. These walls are monoatomic steps with a height equal to one-half of the lattice parameter. Some of them are not closed, suggesting the presence of dislocations going through the surface and bounding the magnetic fault defined by the domain walls. These dislocations have \textit{a priori} a $1/2\,\hkl<111>$ Burgers vector.

In addition to these $1/2\,\hkl<111>$ dislocations, TEM observations reveal that dislocations with \hkl<100> Burgers vectors, also gliding in \hkl{110} planes, are present in magnetically ordered Cr \cite{mclaren1964,reid1966,hale1969}. Reid and Gilbert reported at 300\,K a cross-slip event incompatible with a \hkl<111> screw orientation and requiring a \hkl<100> Burgers vector. Such a Burgers vector was confirmed by Hale and Henderson Brown \cite{hale1969}: using extinction contrast in TEM, they obtained a much higher proportion of \hkl<100> dislocations in Cr than in iron, where these \hkl<100> can result, like in any BCC metal, from junctions between $1/2\,\hkl<111>$ dislocations. Reid \cite{reid1966} showed that, as a consequence of the strong elastic anisotropy of BCC Cr, \hkl<100> dislocations have indeed a similar elastic energy as $1/2\,\hkl<111>$ despite their larger Burgers vectors. Cr magnetism at low temperature should also favor these \hkl<100> dislocations: as \hkl<100> is a periodicity vector of the magnetic order below the Néel temperature, such dislocations can exist without a magnetic fault, in contrast to $1/2\,\hkl<111>$ dislocations. Although these \hkl<100> dislocations may be as important as $1/2\,\hkl<111>$ dislocations in BCC Cr, not much is known about them.

The object of this work is to qualify by means of \textit{ab initio} calculations the influence of magnetism on the plastic behavior of BCC Cr below its Néel temperature. Particularly, we study the competition between dislocations with Burgers vectors $1/2\,\hkl<111>$ and \hkl<100>, which have been both observed experimentally, and discuss the consequences of magnetism on the properties and mobility of the two slip systems. We begin by introducing the general methods used for this study, covering magnetic and elastic properties, and then the generalized stacking fault energies before the two types of screw dislocations, $1/2\,\hkl<111>$ and \hkl<100>, and discussing the obtained results.

\section{Methods and elementary properties}
\subsection{Computational details}
All calculations in the present work are carried out within density functional theory (DFT) as implemented in the \textsc{Vasp} code \cite{kresse1996}. The Kohn-Sham states are represented using a plane-wave basis with a 500\,eV cutoff energy. A projector augmented wave (PAW) potential \cite{blochl1994} is used for Cr including $12$ valence electrons, and the exchange-correlation potential is approximated with the GGA-PBE functional \cite{perdew1996}. The Methfessel-Paxton broadening scheme is used, with a 0.1\,eV width. {A $\Gamma$-centered $k$-point sampling of the Brillouin zone is generated using the Monkhorst-Pack scheme, with a density of 20\,$k$-points per lattice parameter unit length in every direction for each simulation cell.} To investigate the influence of magnetism on the plastic behavior of Cr, we compute all properties in both the non-magnetic (NM) and the antiferromagnetic (AF) phase. As the AF phase is not the true ground state of Cr, we first check that it is a good approximate of the experimental spin density wave which should be more stable below the Néel temperature, as it will be discussed below. Magnetism is treated as collinear within spin-polarized DFT. All relaxation calculations are carried out with fixed periodicity vectors at the equilibrium lattice parameter determined by minimizing the energy of the BCC unit-cell as a function of the lattice parameter in a given magnetic state. The stopping condition is that the remaining forces are less than {5\,meV/{\AA}} on all atoms along all Cartesian directions. In the following, we check that this \textit{ab initio} modeling of Cr gives a good representation of the competition between its magnetic phases, their lattice parameters and their elastic behaviors before moving on to more complex properties controlling plasticity.
\subsection{Stability of magnetic phases}
\begin{figure}[htb]
        \centering
        \includegraphics[width=\linewidth]{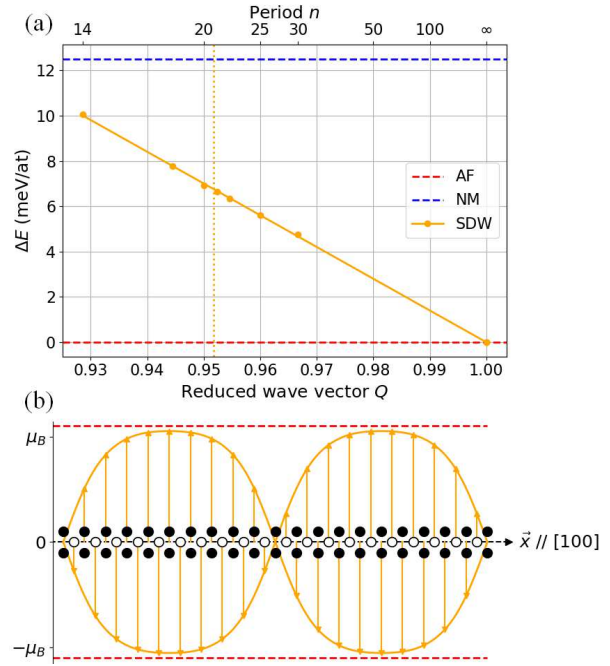}
        \caption{(a) Energy difference $\Delta E$ per atom between the SDW and AF phases as a function of the period $n$ of the SDW and its reduced wave vector $Q=1-1/n$ corresponding to the wave vector $\vec{q}=Q\times2\pi/a_0\hkl[100]$. The horizontal blue and red dashed lines indicate the energy difference for the NM and AF phases respectively. The vertical orange line marks the experimental wave-vector of the SDW. (b) Variation of the magnetic moments along the propagation direction of the SDW for $n=20$. The red lines indicate the magnetic moment of the AF phase. The black and white circles represent corner and body-center atoms of the BCC unit-cell.}
        \label{fig:qSDW}
\end{figure}

Neutron diffraction \cite{fawcett1988} and coherent X-ray diffraction \cite{jacques2009} experiments showed the magnetic ground-state of BCC Cr to be a spin-density wave (SDW) with incommensurate period regarding the crystal lattice below its Néel temperature of 311\,K. The SDW corresponds to a quasi-sinusoidal modulation of the magnitude of the magnetic moments along the propagation of the wave, keeping a locally antiferromagnetic order (Fig. \ref{fig:qSDW}.b). More precisely, from 0 to 123\,K, the SDW is longitudinally polarized with magnetic moments oriented along a \hkl<100> axis of the crystal lattice, roughly corresponding to a period of 20 lattice parameters \cite{fawcett1988}. At 123\,K, the polarization of the SDW switches to {transverse} with {magnetic} moments directed perpendicular to its propagation direction, before vanishing at the Néel temperature. In the following, we examine more closely the stability of the different magnetic phases of Cr.

We consider the three following magnetic phases: NM, AF and SDW. Before discussing the relative stability of the magnetically ordered AF and SDW phases of Cr, it is worth noting that the NM phase is found to have the highest energy among the three (Tab. \ref{tab:elastic}). However, DFT calculations fail to predict the SDW phase as the ground state and invariably find the AF phase as more stable at 0\,K whatever the exchange and correlation functional and the DFT approximations \cite{vanhoof2009,soulairol2010,hafner2002,soulairol2011,cottenier2002}. {Indeed, all Vanhoof \etal \cite{vanhoof2009} using LDA+U, Soulairol \etal \cite{soulairol2010} using both LDA, GGA and mixed LDA-GGA functionals, and Cottenier \etal \cite{cottenier2002} using the FLAPW method with GGA functional, found the SDW to have a higher energy than the AF phase.} 

{All three considered magnetic phases in this work are collinear, the SDW corresponding to a modulation of the spin magnitude, keeping locally an antiferromagnetic order. Neutron diffraction experiments performed on pure bulk Cr \cite{fawcett1988} report no evidence of non-collinear magnetic structures, and confirmed the collinearity of the SDW gound state of Cr. The theoretical work of Soulairol \etal \cite{soulairol2010,soulairol2011} on the relative stability of the magnetic phases of Cr revealed the non-collinear spin spirals states to be unstable for any orientation. This is also reported in the work of Shallcross \etal \cite{shallcross2005} within the KKR scheme. These observations motivated the use of the collinear magnetism approximation in the present work.} Due to finite size of simulation cells, we are not able to consider the incommensurate SDW found experimentally. We have to study commensurate structures with periodicity $n$ equal to an integer number of the lattice parameter $a_0$. In the following, we note $\vec{q}=2\pi/a_0[1-1/n;0;0]$ the wave vector of the SDW. To distinguish between the longitudinal and transverse SDW, spin-orbit coupling has to be taken into account. Soulairol {\etal} \cite{soulairol2010} showed that the energy difference between the two wave polarizations is not relevant with respect to DFT uncertainty. Hence we do not consider SDW polarization. The energy difference per atom between the SDW and the AF phases is presented in Figure \ref{fig:qSDW}.a as a function of the period $n$ of the wave.

In agreement with previous \textit{ab initio} studies \cite{soulairol2010,vanhoof2009,hafner2002}, our calculations lead to a higher energy for the SDW than the AF phase for any period $n$. Its excess energy with respect to the AF phase varies linearly with the wave vector magnitude $1/n$. This discrepancy with experiment is often attributed to the inner limitations of DFT \cite{soulairol2010,hafner2002}. However, Vanhoof \etal \cite{vanhoof2009} offers another explanation. Their approach suggests that the stabilization of the SDW comes from the perturbation of the AF order by the introduction of nodons corresponding to locally zero magnetic moments and the associated entropy. The linear behavior of the energy difference between the SDW and AF phases as a function of $\vec{q}$ corresponds to a nodon excitation energy of 140\,meV, in good agreement with the 152\,meV nodon energy obtained by Vanhoof \etal \cite{vanhoof2009}. 

The variation of the magnetic moments $\mu_i$ along the propagation direction of the SDW takes the form of a Fourier series with only two harmonics
\begin{equation}
\mu_i=M_1\sin{(\vec{q} \cdot \vec{R_i})}+M_3\sin{(3 \vec{q} \cdot \vec{R_i})}+... \label{eq:formeSDW}
\end{equation}
where $M_j$ is the amplitude of the $j$-th term of the Fourier series and $\vec{R_i}$ is the position of the $i$-th atom along the \hkl<100> propagation direction of the SDW. The shape of the wave (Fig. \ref{fig:qSDW}.b) is determined by the two $M_1$ and $M_3$ amplitudes, for which we find $M_1=1.18$ and $M_3=0.15$\,$\mu_B$ for $n=20$. This results in a peak magnetic moment of 1.05\,$\mu_B$, very close to the AF magnitude of 1.1\,$\mu_B$. However, these values are twice the experimental one of approximately 0.5\,$\mu_B$ at 4.2\,K \cite{fawcett1988}. This overestimation of the magnetic moments is a well known discrepancy of the GGA-PBE exchange and correlation functional with experiments and is reported in various DFT studies on the stability of the magnetic phases of Cr \cite{vanhoof2009,soulairol2010,hafner2002}. We find lower lattice parameters for the three phases than the experimental value of 2.884\,{\AA} at 4.2\,K. Still, the equilibrium lattice parameters of the magnetic AF and SDW phases are closer to the experimental value than the NM case as reported in Table \ref{tab:elastic}. The LDA functional gives a better estimation of magnetic moments, however the obtained equilibrium lattice parameters deviate more from the experimental value \cite{soulairol2010}.

In the following, we will use the AF phase as an approximate of the true magnetic state of Cr at low temperature. This choice is motivated by the impossibility to introduce both a spin density wave and a structural defect like a stacking fault or a dislocation in a simulation cell with a reasonable number of atoms. Besides, following the nodon model of Vanhoof \etal \cite{vanhoof2009}, the SDW appears as a perturbation of the AF phase, which may justify the validity of our approximate description of the magnetic order of Cr below the Néel temperature. Finally, Bacon and Cowlam \cite{bacon1969} and Williams and Street \cite{williams1981} have shown that the AF phase becomes more stable than the SDW above roughly 200\,K in strained samples containing dislocations, with the Néel temperature of the AF phase going up to 450\,K. It appears therefore fully legitimate to study dislocation properties in this AF phase.

\subsection{Elastic properties}
We then evaluate the elastic constants of the three considered magnetic phases (NM, AF and SDW). The results are shown in Table \ref{tab:elastic}. The SDW structure has a tetragonal symmetry corresponding to 6 elastic constants, but its anisotropy is very small, with a maximum discrepancy of 6\,GPa between $C_{11}$ and $C_{22}$. The presented results in cubic symmetry are obtained by averaging over the three \hkl[100], \hkl[010] and \hkl[001] wave directions for a SDW with periodicity $n=20$. We note that the obtained values for the AF and SDW magnetic phases are closer to the experimental data at 4.2\,K of Palmer and Lee \cite{palmer1971} than the NM phase, particularly regarding the bulk modulus $B$. Indeed, magnetism is very sensible to volume variation, showing its significant impact on the elastic properties of Cr at low temperature.

\begin{center}
        \begin{table}[htb]
                \caption{Lattice parameter $a_0$ ({\AA}), bulk modulus $B$ and shear moduli $C'=(C_{11}-C_{12})/2$ and $C_{44}$ (GPa), elastic anisotropy ratio $A=C_{44}/C'$ and energy difference $\Delta E$ (meV/atom) with respect to the AF ground-state of the NM, AF and SDW (for a $n=20a_0$ periodicity) magnetic phases of BCC Cr. The experimental data at 4.2\,K are taken from Palmer and Lee \cite{palmer1971}, corresponding to the incommensurate longitudinal SDW.}
                \label{tab:elastic}
                \centering
                \begin{tabular}{c c c c c c c}
                        \hline
                        & $a_0$ & $B$ & $C'$ & $C_{44}$ & $A$ & $\Delta E$ \\
                        \hline
                        NM & 2.847 & 262 & 166 & 98 & 0.59 & 12.5\\
                        AF & 2.865 & 186 & 185 & 96 & 0.52 & 0.0\\
                        SDW & 2.857 & 198 & 187 & 101 & 0.54 & 6.4\\
                        Exp. & 2.884 & 190 & 153 & 104 & 0.68 & / \\
                        \hline
                \end{tabular}
        \end{table}
\end{center}

The elastic constants of the AF and SDW phases are very close, and the differences with experimental data mostly come from an overestimation of $C'$. Most importantly, the shear moduli $C'$ and $C_{44}$ of these two phases are identical within DFT accuracy. As screw dislocations do not induce a variation of volume, they should have almost identical elastic behaviors in both magnetic phases. This comforts us in approximating the low temperature experimental SDW ground state by the AF phase in further calculations such as stacking faults and dislocation properties, which will be also computed in the NM phase.

\section{Generalized stacking faults}
\label{section:gamma}
Before introducing dislocations in the crystal, studying the generalized stacking faults (GSF) \cite{vitek1968} can help to get useful information about the ease to shear the crystal in different planes. GSFs describe the excess energy per unit surface associated with the rigid shearing of the perfect crystal into two halves by a fault vector $\vec{f}$ lying in a given crystallographic plane. The positions of the atoms are allowed to relax only perpendicularly to the plane considered to maintain the fault during relaxation. The map of the relaxed energies as a function of the fault vector $\vec{f}$ is called the $\gamma$-surface.

\subsection{Simulation setup}
We use periodic stackings of crystallographic planes and the shift in atomic positions by the fault vector is applied to the periodicity vector perpendicular to the plane to introduce only one fault per cell and avoid free surfaces. As the dislocations mainly move in the \hkl{110} planes, the full $\gamma$-surface will be studied only for the \hkl{110} planes, but we also consider the projection of the \hkl{112} and \hkl{123} $\gamma$-surfaces on a \hkl<111> direction, and the projection of the \hkl{100} $\gamma$-surface on a \hkl<100> direction. We checked the convergence of the GSF energies as a function of the separation distance between two faulted planes and chose the parameters presented in Table \ref{tab:setupGSF} for the simulation cells.

\begin{center}
        \begin{table}[htb]
                \centering
                \caption{Geometry of the simulation cells used for the GSF calculations in different crystallographic planes. The number of stacked planes $n_Z$ corresponds to a separation distance $d_{fault}$ between two periodic images of the fault.}
                \begin{tabular}{c c c c c c}
                        \hline
                        Plane & $X$ & $Y$ & $Z$ & $n_Z$ & $d_{fault}$\\
                        \hline
                        \hkl{110} & \hkl[11-2] & \hkl[111] & \hkl[-110] & $12$ & $6a_0\sqrt{2}$\\
                        \hkl{112} & \hkl[-110] & \hkl[111] & \hkl[11-2] & $24$ & $4a_0\sqrt{6}$\\
                        \hkl{100} & \hkl[100] & \hkl[010] & \hkl[001] & $40$ & $20a_0$\\
                        \hkl{123} & \hkl[11-1] & \hkl[-54-1] & \hkl[123] & $28$ & $2a_0\sqrt{14}$\\
                        \hline
                \end{tabular}
                \label{tab:setupGSF}
        \end{table}
\end{center}

\subsection{\hkl<111> slip mode}
\label{subsection:slip111}
\begin{figure}[htb]
        \centering
        \includegraphics[width=\linewidth]{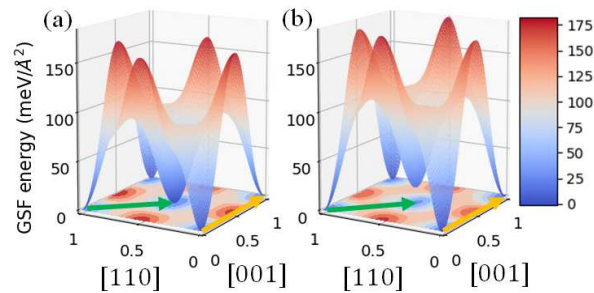}
        \caption{BCC Cr \hkl{110} $\gamma$-surfaces: (a) NM phase, (b) AF phase, showing a magnetic fault at the Burgers vector $\vec{b}=1/2\,\hkl[111]$ indicated by the green arrows. The orange arrows indicate $\vec{b}=\hkl[001]$.}
        \label{fig:surfaces110}
\end{figure}

$1/2\,\hkl<111>$ dislocations in BCC metals glide in one of the three planes with the largest interplanar distance, \hkl{110}, \hkl{112} and \hkl{123}, with a prevalence for \hkl{110} \cite{weinberger2013}, which is also observed in the case of Cr \cite{reid1966,mclaren1964}. The \hkl{110} $\gamma$-surfaces for the NM and AF phases are presented  in Figure \ref{fig:surfaces110} with a sampling of $10$ points per direction and a Fourier series interpolation.

\begin{figure*}[htb]
        \centering
        \includegraphics[width=0.9\linewidth]{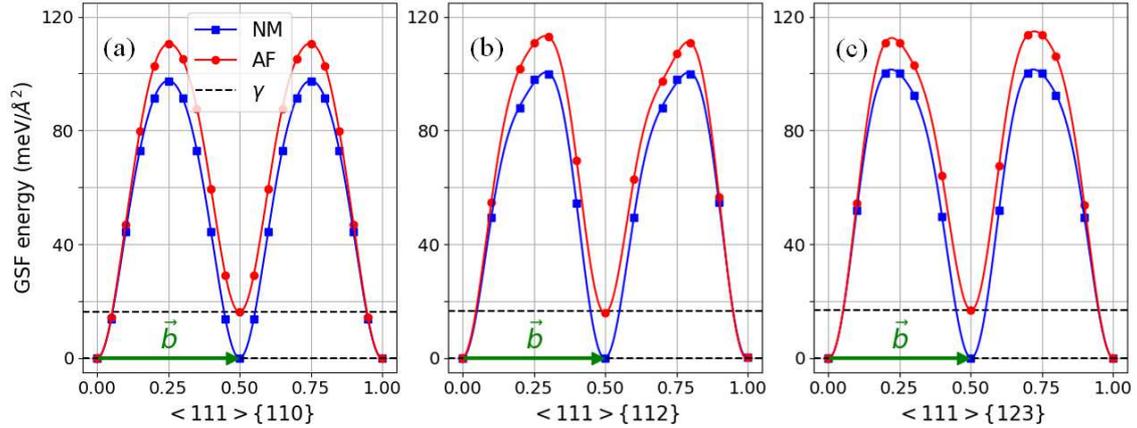}
        \caption{Generalized stacking fault energy along the \hkl[111] direction for a (a) \hkl{110}, (b) \hkl{112}, and (c) \hkl{123} fault plane. The blue squares correspond to the NM phase and the red circles to the AF phase. The green arrows show the Burgers vector $\vec{b}=1/2\hkl[111]$.}
        \label{fig:lignes111}
\end{figure*}

The shape of the $\gamma$-surfaces are very similar in both magnetic phases, except for the introduction of a fault corresponding to the Burgers vector $1/2\,\hkl[111]$ in the AF phase. This vector is a periodicity vector of the BCC lattice and there is no excess energy in the NM phase as the perfect BCC lattice is recovered. But $\vec{b}=1/2\,\hkl[111]$ breaks the AF magnetic order of Cr, thus leading to a magnetic fault in the AF phase. The minimum corresponding to this magnetic fault is better visualized on the $\gamma$-line defined as the projection of this $\gamma$-surface in the \hkl<111> direction (Fig. \ref{fig:lignes111}a). The same \hkl<111> $\gamma$-lines have been also calculated for the \hkl{112} and \hkl{123} fault planes (Fig. \ref{fig:lignes111}b and c respectively), which exhibit the well known twinning / anti-twinning (T/AT) asymmetry \cite{rodney2017}.

A magnetic fault is obtained in all three planes for a fault vector equal to $\vec{b}=1/2\,\hkl[111]$. These magnetic faults have very close energies per surface unit: $\gamma_{110}=16.2$, $\gamma_{112}=16.4$, and $\gamma_{123}=16.7$\,meV/{\AA$^2$}. These values are obtained after full relaxation along all three Cartesian axis to check the stability of the fault. The magnetic fault results from the shearing of the crystal forcing two parallel spins to face each other creating a magnetic frustration partially resolved by locally reducing the magnitude of the atom magnetic moments. This can also be regarded as an antiphase magnetic domain wall separating two reversed-magnetization half crystals. {Another possibility to partially resolve this magnetic frustration would be by rotating the magnetic moments in the vicinity of the fault plane. We checked if this configuration was also possible considering the \hkl{110} magnetic fault as an example by relaxing the system taking account of non-collinear magnetism and spin-orbit coupling. We initialize all magnetic moments along the $X$ axis of the simulation cell, which is perpendicular to the normal of the fault plane $Z$, except for the two closest planes from the fault, where they are initialized along the $Z$ axis to insert a non-collinear perturbation. This initial non-collinear magnetic structure was found to relax to the same as the collinear one.}

Except for this excess energy at the center of the $\gamma$-lines, their shapes are very similar in the NM and AF phases in all three considered fault planes, indicating a weak influence of magnetism on the relative ease to shear these planes. Also, regardless of the magnetic phase, the peak amplitude and slope of the three \hkl<111> $\gamma$-lines are almost identical in all three fault planes. Hence, no particular slip system appears to be easier to activate than any other, and \hkl{110} does not seem to be the easiest one to shear even if it is the main experimental slip plane. This shows a fine description of the structure and mobility of dislocations is required to have a good understanding of the mechanisms involved in the plastic deformation of Cr.

\subsection{\hkl<100> slip mode}
\label{subsection:slip100}
\begin{figure}[htb]
        \centering
        \includegraphics[width=\linewidth]{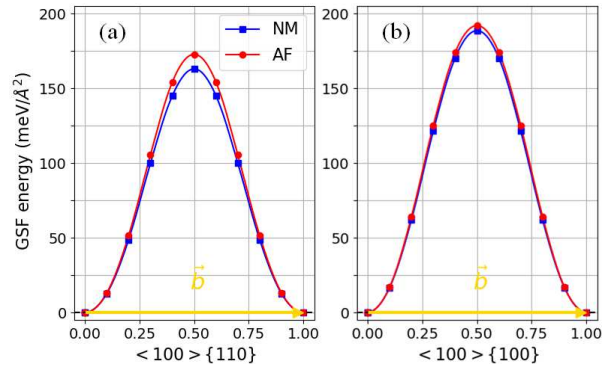}
        \caption{Generalized stacking fault energy along the \hkl[100] direction for a (a) \hkl{110}, and (b) \hkl{100} fault plane. The blue squares correspond to the NM phase and the red circles to the AF phase. The gold arrows show the Burgers vector $\vec{b}=\hkl[100]$.}
        \label{fig:lignes100}
\end{figure}

As we are also interested in \hkl<100> screw dislocations, it is interesting to look at the generalized stacking fault in the crystallographic planes containing this \hkl<100> direction, \textit{i.e.} \hkl{110} and \hkl{100} planes. Figure \ref{fig:lignes100} shows the resulting $\gamma$-lines corresponding to this \hkl[100] direction for these two planes.

The \hkl<100> $\gamma$-line have a lower energy maximum in \hkl{110} planes than in \hkl{100} planes in both NM and AF phases, with no magnetic fault as expected from the magnetic order of Cr. This suggests \textit{a priori} an easier glide of \hkl<100> dislocations in \hkl{110} planes as observed experimentally \cite{reid1966,hale1969}. However, the shearing of the crystal in any direction still induces a minor local frustration of the atom magnetic moments resulting in higher fault energies in the AF phase, as also observed along \hkl<111> directions (Fig. \ref{fig:lignes111}).

\section{$1/2\,\hkl<111>$ screw dislocation}
As observed in the previous section, stacking fault energies are only a first step to rationalize Cr plasticity. A more accurate understanding requires an atomic description of screw dislocation cores.
\subsection{Simulation setup}
The geometry of the simulation supercells used for the study of $1/2\,\hkl<111>$ screw dislocations have Cartesian directions such that the \hkl{110} glide plane of the dislocations is oriented with its normal along $Y \parallel \vec{u}_2=\hkl[-110]$ axis, with the glide direction along $X \parallel \vec{u}_1=\hkl[11-2]$ axis, and the dislocation line along the $Z \parallel \vec{u}_3=1/2\hkl[111]$ axis. The periodicity vectors $(\vec{p}_1,\vec{p}_2,\vec{p}_3)$ of the supercells are represented by 
\begin{eqnarray}
    \begin{array}{c}
    \vec{p}_1=\lambda_1\vec{u}_1-\lambda_2\vec{u}_2+\lambda_3\vec{u}_3\\ \label{eq:lambda}
    \vec{p}_2=\lambda_1\vec{u}_1+\lambda_2\vec{u}_2+\lambda_3\vec{u}_3\\
    \vec{p}_3=\vec{u}_3\\
    \end{array}
\end{eqnarray}
with the values of $\lambda_i$ for the different cell sizes recapitulated in Table \ref{tab:setupdislo}. We use a quadrupolar array of dislocation dipoles with three-dimensional periodic boundary conditions \cite{rodney2017,clouet2018} to limit the elastic interactions between periodic images. In this setup, the two dislocations of a same dipole are separated from each other by a vector $(\vec{p}_1+\vec{p}_2)/2$ when aligned horizontally, or $(\vec{p}_1-\vec{p}_2)/2$ when aligned vertically. For the NM phase, the supercell is $1b$-high, with $b=a_0\sqrt{3}/2$, whereas for the AF phase we need to use a $2b$-high supercell as the Burgers vector is not a periodicity vector of the AF magnetic order. The dislocations are introduced in the simulation cells using anisotropic elasticity theory taking full account of periodicity, with a homogeneous strain applied to the lattice vectors of the cell to cancel the plastic strain created by the dislocation dipole \cite{rodney2017,clouet2018}. Atomic positions are then fully relaxed with fixed periodicity vectors.

\begin{center}
        \begin{table}[htb]
                \centering
                \caption{Parameters $\lambda_i$ defining the periodicity vectors of the supercells used for the study of the $1/2\,\hkl<111>$ screw dislocation (\ref{eq:lambda}) corresponding to a number of atoms $N$ for a $1b$-high supercell.}
                \begin{tabular}{c c c c}
                        \hline
                        $N$ & $\lambda_1$ & $\lambda_2$ & $\lambda_3$\\
                        \hline
                        $135$ & $5/2$ & $9/2$ & $0$\\
                        $187$ & $17/6$ & $11/2$ & $2/3$\\
                        $209$ & $19/6$ & $11/2$ & $1/3$\\
                        $273$ & $7/2$ & $13/2$ & $0$\\
                        \hline
                \end{tabular}
                \label{tab:setupdislo}
        \end{table}
\end{center}

\subsection{Core structure}
\begin{figure}[htb]
        \centering
        \includegraphics[width=0.9\linewidth]{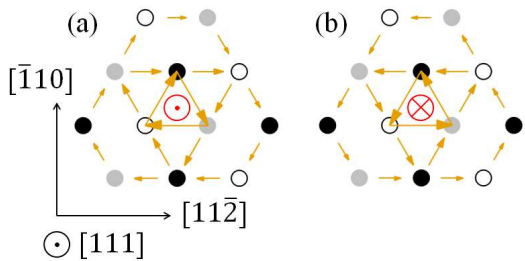}
        \caption{Differential displacements map showing the core structure 
        of the $1/2\,\hkl<111>$ screw dislocation in the NM phase in (a) easy configuration, (b) hard configuration. The atoms are represented by different symbols according to their height along \hkl[111]. An arrow joining two atoms corresponds to a differential displacement of $b/3$ along \hkl[111], with $b=a_0\sqrt{3}/2$ the norm of the Burgers vector.}
        \label{fig:coresNM}
\end{figure}

The structures obtained for the $1/2\,\hkl<111>$ screw dislocation after atomic relaxation can be visualized using differential displacement maps along the \hkl<111> direction as introduced by Vitek \cite{vitek1970} and presented in Figure \ref{fig:coresNM} for the NM phase.

\begin{figure}[htb]
        \centering
        \includegraphics[width=\linewidth]{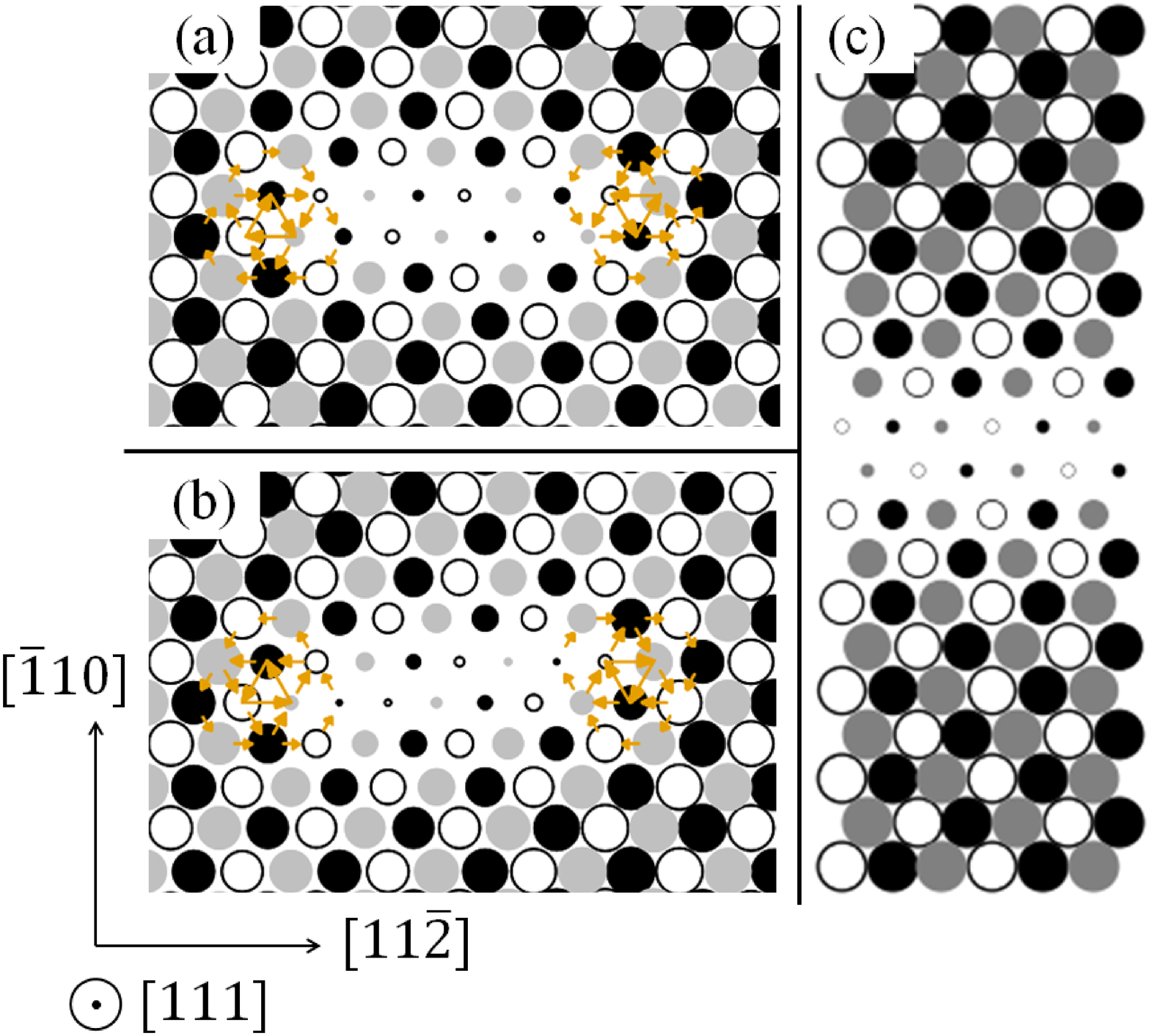}
        \caption{Differential displacements map showing the core structure 
        of the $1/2\,\hkl<111>$ screw dislocation in the AF phase with the magnetic fault located along a \hkl{110}
        plane (a) easy configuration, (b) hard configuration. (c) Magnetic fault in a \hkl{110} plane for comparison. The diameter of the circles represents the amplitude of the magnetic moments on each atomic site. Two touching circles corresponding to the bulk DFT value of 1.1\,$\mu_B$. {The smallest circles correspond to a magnetic moment of 0.2\,$\mu_B$.}}
        \label{fig:cores110}
\end{figure}

\begin{figure}[htb]
        \centering
        \includegraphics[width=0.9\linewidth]{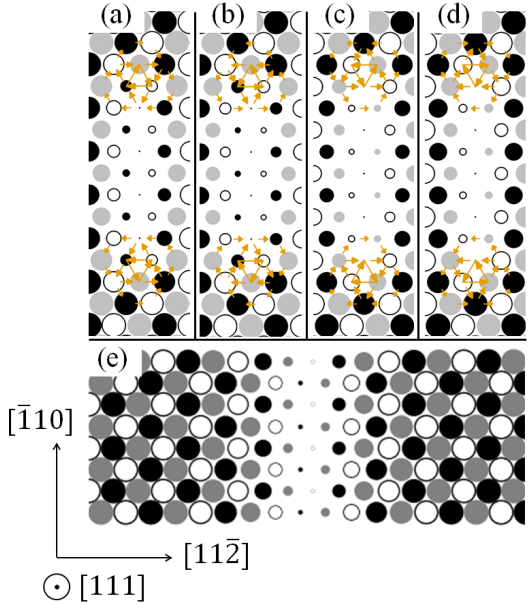}
        \caption{Differential displacements map showing the core structure 
        of the $1/2\,\hkl<111>$ screw dislocation in the AF phase with the magnetic fault located along a \hkl{112}
        plane: (a) and (b) easy and hard configurations for the up triangles; (c) and (d) for the down triangles. (e) Magnetic fault in a \hkl{112} plane for comparison. {The smallest circles correspond to zero magnetic moment.}}
        \label{fig:cores112}
\end{figure}

The core structure is shown in two configurations: the easy core, which is the ground-state (Fig. \ref{fig:coresNM}.a), and the hard core which is an unstable maximum (Fig. \ref{fig:coresNM}.b). Both configuration have a compact core as observed in other BCC metals using DFT calculations \cite{rodney2017}. The easy core displays reversed helicity of the three \hkl<111> atomic columns in the vicinity of the dislocation center, whereas the hard core constrains the three columns to be at the same height. To relax the unstable hard core structure, the coordinates of the atoms along the $Z$ axis are frozen for the three nearest atomic columns and all other atoms are allowed to fully relax in the three Cartesian directions. Differential displacement maps for the AF phase are presented in Figures \ref{fig:cores110} and \ref{fig:cores112}.

The dislocation core structure in both easy and hard configurations are the same in the NM and AF phases. The only difference is the magnetic fault between two dislocations of the same dipole due to the Burgers vector not being a periodicity vector of the AF magnetic order. This fault {appears in the region between the dislocations which has been sheared by the Burgers vector to create the dipole, starting from a perfect crystal.} It gives rise to a magnetic frustration resolved by reducing the magnitude of the atom magnetic moments in the vicinity of the fault plane. A different type of representation is adopted to better visualize this fault where the diameter of the represented atoms is proportional to its magnetic moment (Figures \ref{fig:cores110} and \ref{fig:cores112}). The magnetic fault is located in either a \hkl{110} or \hkl{112} plane depending on initial choice for the vector joining the two dislocations of the dipole, $(\vec{p}_1+\vec{p}_2)/2$ or $(\vec{p}_1-\vec{p}_2)/2$ respectively. 

The center of a $1/2\,\hkl<111>$ screw dislocation in its easy or hard configuration is located at the center of gravity of triangles formed by three adjacent atomic columns along a \hkl<111> direction (Fig. \ref{fig:coresNM}.a). Dislocations with $+b$ Burgers vector are located at the center of triangles pointing up, and $-b$ dislocations of triangles pointing down. Hence, the length of the vector joining two dislocation centers in the $Y$ direction varies by a small amount $\pm \delta=a_0\sqrt{2}/6$ if it links triangles pointing up and down or down and up, and depending on the core configurations, easy or hard. This is of no importance for the NM and AF cases when the fault lies in a \hkl{110} plane, but it changes the structure and length of the \hkl{112} magnetic fault (Fig. \ref{fig:cores112}).

We also show the structure of the infinite faults as given by the local minima in the GSFs (Fig. \ref{fig:lignes111}) in Figures \ref{fig:cores110}.c and \ref{fig:cores112}.e in \hkl{110} and \hkl{112} planes respectively. The structure of the infinite fault is identical as the one observed for a dislocation dipole laying in a \hkl{110} plane, but slightly differs for a \hkl{112} plane. In the GSFs, the fault lies between two adjacent \hkl{112} planes so that no magnetic moment is strictly zero. For the dislocation dipoles, the fault is located on a \hkl{112} atomic plane resulting in exactly zero magnetic moments in that plane. Otherwise, the structure of the magnetic fault is nearly identical for both the easy and the hard core configurations, regardless of the orientation of its plane or the setup.

\subsection{Core energies}
\label{subsection:ecore111}
The total energy $E^{tot}$ of the simulation cells can be partitioned as 
\begin{equation}
E^{tot}=E^{bulk}+E^{elastic}+2E^{core}+E^{fault}, \label{eq:etot}
\end{equation}
where $E^{bulk}$ is the energy of the perfect unfaulted BCC lattice, $E^{elastic}$ is the elastic energy of the dislocation dipole including the interaction between periodic images, $E^{core}$ is the core energy of the dislocations, and $E^{fault}$ is the energy of the magnetic fault. All energies are normalized by the length of the simulation cell along the $Z$ axis to account for the different cell heights between the NM and AF phases. The elastic energy of the dipole is evaluated using anisotropic elasticity theory with the \textsc{BABEL} package \cite{babel} by defining a core radius $r_c=b=a_0\sqrt{3}/2$. 

\begin{figure}[htb]
        \centering
        \includegraphics[width=\linewidth]{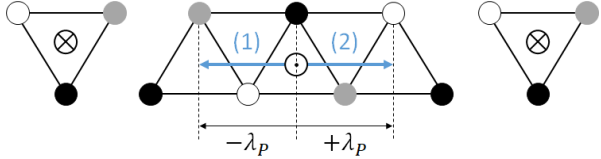}
        \caption{Schematic of the dislocation arrangements to calculate the energy of the magnetic fault separating two dislocations of the same dipole. The $-b$ dislocation is fixed while the $+b$ dislocation is moved one Peierls valley on the left (1) and on the right (2), with $\lambda_P=a_0\sqrt{2/3}$ the distance between two adjacent Peierls valleys.}
        \label{fig:fault}
\end{figure}

We first look at the contribution of the magnetic fault which exists in the AF phase. This fault is assumed to have an energy $E^{fault}=\gamma d$, with $\gamma$ the energy of the magnetic fault per surface unit and $d$ the distance between the two dislocations of the dipole. This fault energy should be equal to the one determined in the previous section for an infinite fault plane, which we propose to check now. To evaluate $\gamma$ directly from \textit{ab initio} modeling of dislocations, we use the method sketched in Figure \ref{fig:fault}.

When one of the two dislocations is moved from its initial perfect quadrupolar position to one Peierls valley on the left or on the right while keeping the other fixed, the elastic energy of the dipole is the same as the distance between its periodic images is unchanged. The core energy should also be the same. Therefore, the energy difference between the two configurations is only due to the magnetic fault having different lengths, $\pm \lambda_P$ with respect to the quadrupolar arrangement. The energy of the magnetic fault $\gamma$ can then be expressed directly as $(E_{(1)}-E_{(2)})/4b\lambda_P$. We apply this procedure only for a magnetic fault located in a \hkl{110} plane. This leads to $\gamma=16.3$\,meV/{\AA$^2$}, which perfectly agrees with the value we obtained from GSF, $\gamma_{110}=16.2$\,meV/{\AA$^2$}. The small difference might be due to boundary effects in the vicinity of the dislocation cores. This shows that whether the fault arises from the rigid shearing of the crystal or from the introduction of dislocations, the same phenomenom is involved, at least for \hkl{110} planes. As very similar values were also obtained for infinite \hkl{112} and \hkl{123} faults, $\gamma_{112}=16.4$ and $\gamma_{123}=16.7$\,meV/{\AA$^2$}, the same value of $\gamma=16.3$\,meV/{\AA$^2$} will be used in the following for a dislocation dipole, regardless of the plane of the magnetic fault.

\begin{figure}[htb]
        \centering
        \includegraphics[width=\linewidth]{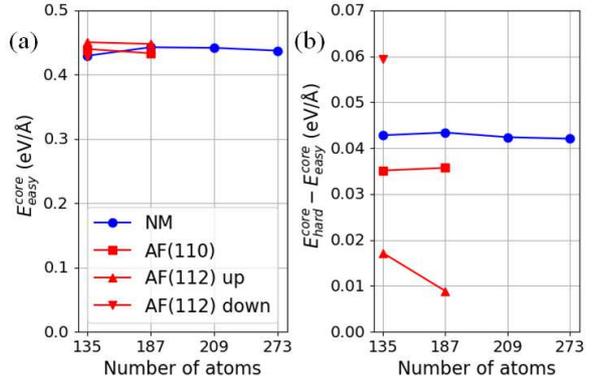}
        \caption{Convergence of the dislocation core energy with respect to the cell size for the NM and AF phases with a magnetic fault located in either a \hkl{110} or \hkl{112} plane. (a) Core energy of the easy core configuration. (b) Energy difference between the hard and easy core configurations. The core radius is $r_c=b$.}
        \label{fig:conv}
\end{figure}

We then check the convergence of the dislocation core energies with respect to the simulation cell size with the parameters of Table \ref{tab:setupdislo}. The results are shown in Figure \ref{fig:conv} for both the NM and AF phases, and for a magnetic fault in the AF phase located in a \hkl{110} and a \hkl{112} plane.

We note that the core energies are almost independent of cell size and of dipole fault plane for the AF phase. This shows that the partition of the total energy proposed in equation \ref{eq:etot} is relevant, with the elastic and magnetic contributions being well evaluated, leading to a core energy independent on dislocation environment. {We did not consider larger simulation cells for the AF phase because the convergence of the core energies is already very good, and the computational cost of these calculations is 8\,times higher than in the NM phase, with twice the number of atoms and the treatment of magnetism.} As expected, the energy of the hard core configuration is higher than the easy core in both magnetic phases. This energy difference shows larger variations in the AF phase when the fault lies in a \hkl{112} plane, with a dependence on the dislocation position either in an up or down triangle (Fig. \ref{fig:conv}.b). As both easy and hard core configurations have the same elastic energy, such a perturbation of the core inevitably arises from an approximate evaluation of the energy contribution of the magnetic fault. This statement is further supported by our previous observation that the structure of the \hkl{112} magnetic fault in the dislocation dipole slightly differs from the infinite fault (Fig. \ref{fig:cores112}). We will therefore prefer for the AF phase the setup with a magnetic fault lying in a \hkl{110} plane in the following, in particular for the calculation of the Peierls energy barrier.

Comparing the results obtained in the NM and AF magnetic phases, one sees that, once the energy contribution of the magnetic fault in the AF phase has been removed, magnetism has only a marginal impact on dislocation energies. Almost the same core energies are obtained in the magnetic phases, with only a slightly smaller energy difference between hard and easy core configurations in the AF than in the NM phase, with 35 instead of 43\,meV/{\AA}.

\subsection{Peierls energy barrier}
\label{sec:Peierls111}
\begin{figure}[htb]
        \centering
        \includegraphics[width=\linewidth]{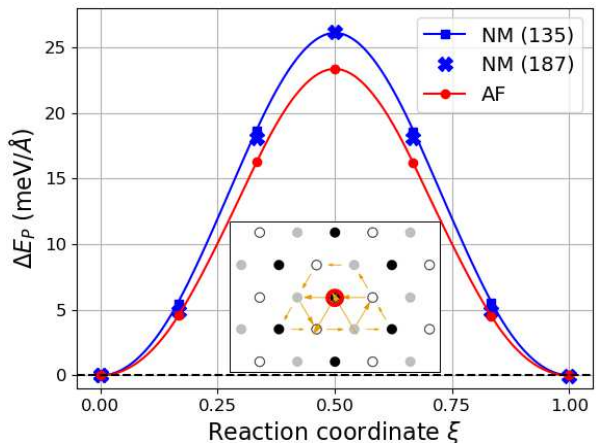}
        \caption{Peierls barrier for a $1/2\,\hkl<111>$ screw dislocation gliding in a \hkl{110} plane for the NM and AF phases. For the AF phase, the magnetic fault is located in the glide plane, \hkl(-110). The inset shows the differential displacement map for the saddle point configuration in the NM phase. The red circled atomic column has a constrained displacement along $Z$ in the AF phase.}
        \label{fig:peierls111}
\end{figure}

We then determine the Peierls energy barrier opposing the $1/2\,\hkl<111>$ screw dislocation glide in a \hkl{110} plane by moving the two dislocations of the dipole in the same direction along $X$ from their initial equilibrium easy configuration to the next nearest, corresponding to the next Peierls valley. This way, the distance between dislocations does not change during the crossing of the barrier so both the elastic and magnetic fault energies remain constant. This is done in the NM and AF phases using the 135-atom cell with height $1b$ and $2b$ respectively. The magnetic fault generated by the dislocation dipole in the AF phase is located in the glide plane \hkl(-110). The minimum energy path is found using the nudged elastic band (NEB) method as implemented in the VASP code. We use five intermediate images between the initial and final states, with a spring constant between images of 0.5\,eV/{\AA}. When performing the calculation in the AF phase, we observed an asynchronous glide of the two dislocations preventing one from ascribing half of the energy variation to each gliding dislocation. To avoid this artefact {found with the unconstrained NEB calculation}, we {performed a second NEB calculation where we} constrained the displacement of the atomic column circled in red in the inset of Figure \ref{fig:peierls111} along the $Z$ axis to be the same, in absolute value, for both $+b$ and $-b$ dislocations. These atomic columns correspond to the most displaced atoms during the crossing of the Peierls barrier. This constraint ensures a synchronous movement of the two dislocations. The resulting barriers are presented in Figure \ref{fig:peierls111}.

The barriers in both magnetic phases have the same shape, with a higher maximum in the NM phase. The calculation in the NM phase was also carried out using the 187-atom supercell and we note a very satisfying convergence of the Peierls barrier with respect to the simulation cell size. The dislocation core structures along the minimum energy path are the same in the two phases, and the magnetic fault \textit{a priori} does not disrupt the structure of the dislocations along the path. The differential displacement map of the saddle point configuration in the NM phase is shown in the inset of Figure \ref{fig:peierls111}, which differs from the hard core configuration as the dislocation drifts away from it during the crossing of the Peierls barrier. Hence the saddle point energy is lower than the energy difference between easy and hard configurations.

\section{\hkl<100> screw dislocation}
\subsection{Simulation setup}
For the study of the \hkl<100> screw dislocation, we use a simulation cell containing 200 atoms for both NM and AF phases with periodicity vectors $\vec{p}_{\,1}=n\hkl[100]$, $\vec{p}_{\,2}=n\hkl[010]$, and $\vec{p}_{\,3}=\hkl[001]$, with $n=10$. The crystal is oriented such that $X \parallel \hkl[-1-10]$, $Y \parallel \hkl[1-10]$, and $Z=\hkl[001]$. The dislocation dipole is introduced in the cell using anisotropic elasticity following a quadrupolar arrangement. This is the only cell size considered, corresponding to a dislocation distance equivalent to the 187-atom cell for the study of the $1/2\,\hkl<111>$ screw dislocation. We infer that this size is large enough to ensure convergence of the properties of interest as the dislocation core structure is also compact.

\subsection{Core structure and energy}
\begin{figure}[htb]
        \centering
        \includegraphics[width=\linewidth]{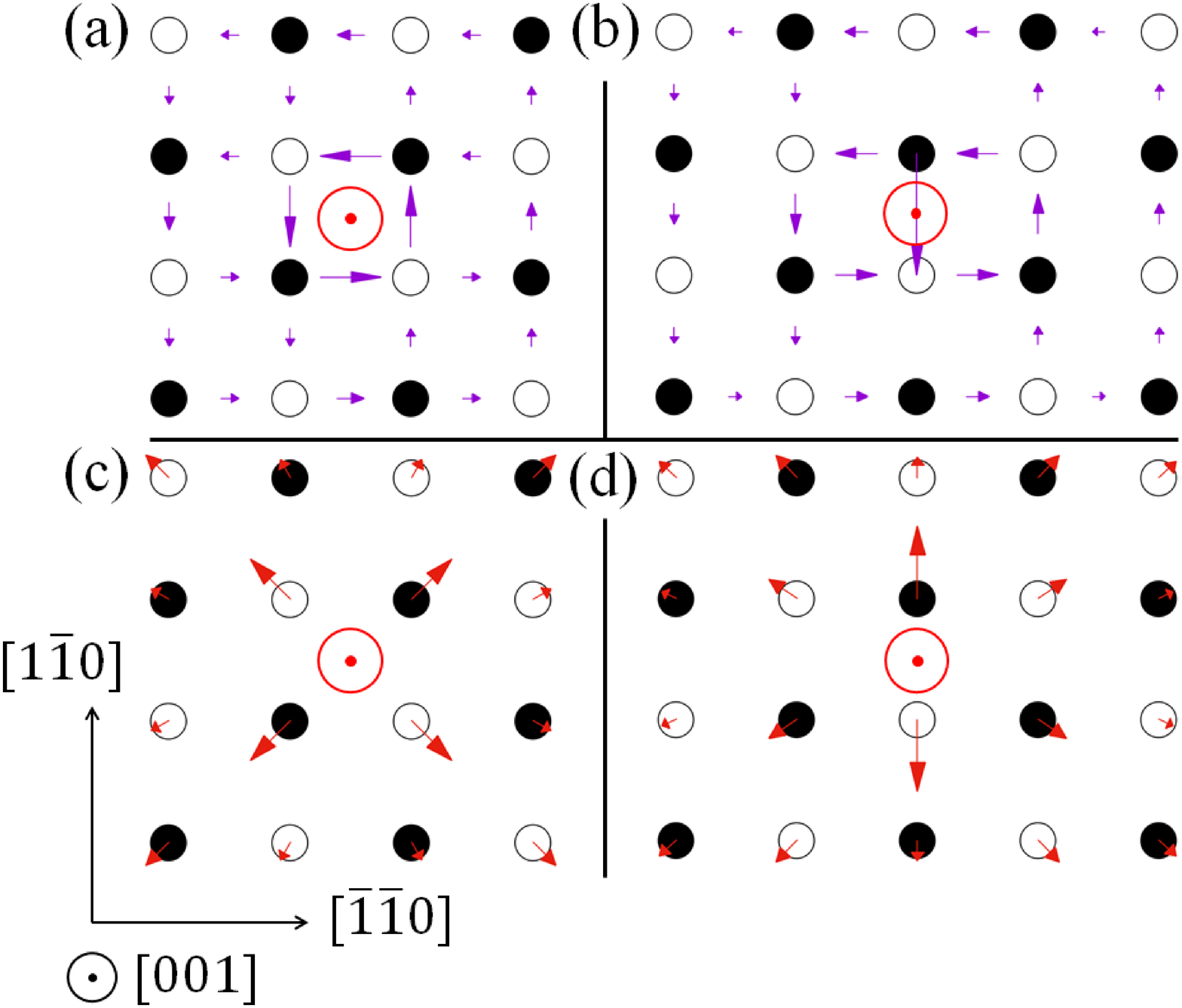}
        \caption{Core structure of the \hkl<100> screw dislocation in the NM phase: (a) and (c) ground state configuration, (b) and (d) metastable configuration. The atoms are represented by different symbols according to their height along \hkl[001]. (a) and (b) show the differential displacement maps. An arrow joining two atoms corresponds to a differential displacement of $b/2$ along \hkl[001], with $b=a_0$ the norm of the \hkl<100> Burgers vector. (c) and (d) show the edge component of the dislocation, \textit{i.e.} the atom displacement projected on the \hkl(001) plane.}
        \label{fig:cores100}
\end{figure}

The dislocation core structure is presented in Figure \ref{fig:cores100} for the NM phase. We consider two positions for the dislocation, one leading to the ground state where the dislocation is located at the center of four \hkl<100> atomic columns along the $Z$ axis, and another one leading to a configuration with higher energy with the dislocation located between two atomic columns along a \hkl{110} plane. We will show in the following that the latter actually coincides with the saddle point configuration when the dislocation is gliding in a \hkl{110} plane.

We find a compact core structure for both configurations, with no spreading. The core structures in the two considered configurations are identical in the NM and AF phases, with no magnetic fault introduced in the system as $\vec{b}=\hkl<100>$ is a periodicity vector of the AF magnetic order. The edge component of the dislocations (Fig \ref{fig:cores100}.c and \ref{fig:cores100}.d) show a slight dilatation in the vicinity of the dislocation center. The core energy of this \hkl<100> screw dislocation is 0.718 and 0.737\,eV/{\AA} in the NM and AF phases respectively, using the same core radius $r_c=a_0\sqrt{3}/2$ as for the $1/2\,\hkl<111>$ dislocation. The energy of the metastable configuration, defined with respect to the ground state, is respectively 25 and 20\,meV/{\AA} in the NM and AF phases.

\subsection{Peierls energy barrier}
\label{sec:Peierls100}
\begin{figure}[htb]
        \centering
        \includegraphics[width=\linewidth]{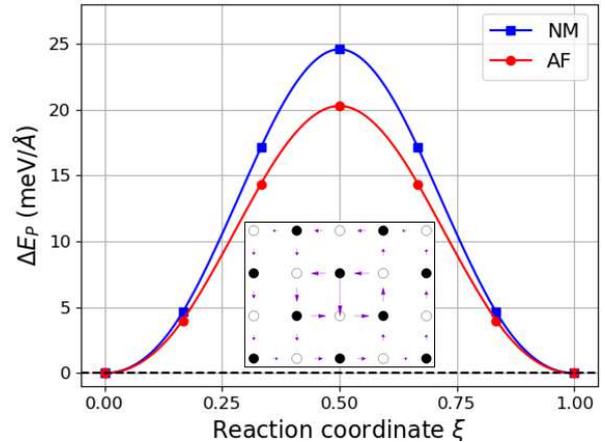}
        \caption{Peierls barrier for a \hkl<100> screw dislocation gliding in a \hkl{110} plane for the NM and AF phases. The inset shows the differential displacement map for the saddle point configuration in the NM phase.}
        \label{fig:peierls100}
\end{figure}

We then determine the Peierls energy barrier opposing the \hkl<100> screw dislocation glide in a \hkl{110} plane using the NEB method by moving the two dislocations of the dipole along the $X$ axis in the same direction from their initial stable core position to the next nearest along the glide direction. The results are shown in Figure \ref{fig:peierls100}. We did not consider glide in \hkl{100} planes as the dislocations would have to cross a \hkl<100> atomic column and no experimental observation report such glide plane.

The barriers have the same shape in both phases, with only a lower energy maximum in the AF phase. The heights of these energy barriers are equal in both cases to the energy difference between the metastable configuration and the ground state identified in the previous section. Along the minimum energy path, the dislocation structures are identical in both phases. We show in the inset of Figure \ref{fig:peierls100} the configuration of the dipole at the saddle point in the NM phase, which corresponds to the metastable configuration of Figure \ref{fig:cores100}.b.

\section{Discussion}
\subsection{Magnetic fault for $1/2\,\hkl<111>$ dislocations}
We begin this section by discussing the consequences of magnetism in the AF phase on the properties and mobility of the $1/2\,\hkl<111>$ screw dislocation. The main effect of magnetism is the existence of a magnetic fault created by the $1/2\,\hkl<111>$ dislocation in its glide plane. This fault exerts a force on the dislocation which needs to be equilibrated by an applied stress. This stress is given by $\tau=\gamma/b$, with $\gamma=16.3$\,meV/{\AA$^2$} the energy of the magnetic fault, leading to a back-stress $\tau=1$\,GPa, for a dislocation gliding in \hkl{110} planes. This stress is too high to allow for the existence of isolated $1/2\,\hkl<111>$ dislocations carrying magnetic faults. Indeed, no such magnetic fault has been reported in TEM observations of bulk BCC Cr strained under its Néel temperature \cite{mclaren1964,reid1966}, in agreement with the associated high energy cost. Such magnetic faults bounded by a dislocation have been observed only on surfaces, with both the magnetic fault and the bounding dislocation ending up at the surface \cite{kleiber2000,ravlic2003}. 

The magnetic fault therefore needs to be closed by another topological defect. This constrains $1/2\,\hkl<111>$ dislocations to coexist pairwise, leading to a super-dislocation dissociated into two partial dislocations separated by a magnetic fault. Considering the different vectors of the $1/2\,\hkl<111>$ family, one obtains such a super-dislocation with \hkl<111>, \hkl<110>, or \hkl<100> Burgers vectors.

\hkl<111> super-dislocations are the result of two partial dislocations with the same $1/2\,\hkl<111>$ Burgers vector, following the reaction $1/2\,\hkl[111]+MF+1/2\,\hkl[111] \rightarrow \hkl[111]$, with $MF$ the magnetic fault. The dissociated configuration corresponding to two $1/2\,\hkl<111>$ dislocations separated by a magnetic fault is energetically more favorable than the single \hkl<111> dislocation, and if there was no magnetic fault as in the NM or the disordered paramagnetic phases, the two partial dislocations would glide apart at an infinite equilibrium distance. The magnetic fault in the AF phase prevents infinite separation. We can evaluate the equilibrium dissociation distance between $1/2\,\hkl<111>$ partial dislocations using elasticity theory, with the following expression for the energy variation arising from the dissociation 
\begin{equation}
\Delta E_{diss}(d)=-b_i^{(1)}K_{ij}b_j^{(2)}\ln{\dfrac{d}{r_c}}+\gamma d, \label{eq:dissociation}
\end{equation}
where $d$ is the dissociation distance, $b^{(1)}$ and $b^{(2)}$ are the Burgers vectors of the two partial dislocations (with $b^{(1)}=b^{(2)}$ in this case), $K$ is the Stroh tensor depending on the elastic constants and on the orientation of the dislocation line vector, $r_c$ is the core radius, and $\gamma$ is the energy of the magnetic fault. The equilibrium dissociation distance $d_{eq}$ is found by minimizing the above expression \ref{eq:dissociation} with respect to $d$, which gives
\begin{equation}
d_{eq}=\dfrac{b_i^{(1)}K_{ij}b_j^{(2)}}{\gamma}. \label{eq:distance}
\end{equation}
For the screw orientation of the \hkl<111> dislocation, an analytic expression can be obtained for the Stroh tensor, leading to
\begin{equation}
d_{eq} = \frac{ \sqrt{ C' \, C_{44} } \, b^2}{2\pi\,\gamma}, \label{eq:distance_screw}
\end{equation}
with $b=a_0\sqrt{3}/2$. Using the values of Table \ref{tab:elastic} for the AF phase and $\gamma=16.3$\,meV/{\AA$^2$} for the magnetic fault energy, we find $d_{eq}=55$\,{\AA} as the equilibrium dissociation distance of the \hkl<111> screw dislocation. Depending on its orientation, the dissociation distance ranges from 54 to 59\,{\AA}, a small variation which is due to the compensation between the effects of dislocation character and of elastic anisotropy of AF Cr. This dissociation distance is short, thus potentially explaining why no TEM observation has reported such a dissociation until now. The total Burgers vector of this dislocation is \hkl<111>, which cannot be easily distinguished experimentally from the usual $1/2\,\hkl<111>$ vector known in BCC metals. TEM observation using the extinction method, \textit{i.e.} $\vec{g}\cdot\vec{b}$ contrast, which concluded to $1/2\,\hkl<111>$ Burgers vector in AF Cr \cite{reid1966,mclaren1964,hale1969} are also compatible with \hkl<111> dislocations. This is also true when the Burgers vector is determined from the screw orientation defined as the intersection of cross-slipped planes.

$1/2\,\hkl<111>$ dislocations can also be combined to form \hkl<100> dislocations, following the scheme $1/2\,\hkl[111]+MF+1/2\,\hkl[1-1-1] \rightarrow \hkl[100]$. The dissociation of \hkl<100> dislocations corresponding to the reverse reaction is unstable: the elastic coefficient $b_i^{(1)}K_{ij}b_j^{(2)}$ appearing in Eq. \ref{eq:distance} is negative for all dislocation characters, both in the NM and AF phases. This agrees with the compact core found for the screw orientation in our \textit{ab initio} calculations. Thus, no magnetic fault is created by dissociation of \hkl<100> dislocations.

The last possibility is the creation of \hkl<110> dislocations, following the reaction $1/2\,\hkl[111]+MF+1/2\,\hkl[11-1] \rightarrow \hkl[110]$. The reverse reaction corresponding to the dissociation of \hkl<110> dislocations into two $1/2\,\hkl<111>$ partial dislocations is energetically more favorable for any orientation, leading to a dissociation distance in the AF phase varying from 12\,{\AA} for the screw orientation to 31\,{\AA} for the edge orientation. However, this \hkl<110> dislocation can also dissociate in two \hkl<100> dislocations, $\hkl[110] \rightarrow  \hkl[100]+\hkl[100]$, without the creation of any magnetic fault as \hkl<100> Burgers vectors are periodicity vector of the magnetic order of the AF phase. This reaction is energetically favorable for dislocation characters between 45{\degree} and edge, showing that \hkl<110> dislocations are unstable for such orientations. This probably explains why no TEM observation has reported the presence of such \hkl<110> dislocations which may only exist as junctions.

\subsection{Competition between $1/2\,\hkl<111>$ and \hkl<100>}
As noted by Reid \cite{reid_anisotropy1966}, \hkl<100> and $1/2\,\hkl<111>$ dislocations have a close elastic energy in Cr as a result of its strong elastic anisotropy. With an anisotropy coefficient $A=C_{44}/C'$ equal to 0.59 and 0.52 respectively in the NM and AF phases, this is true for both magnetic phases. Although our \textit{ab initio} calculations lead to larger core energies for \hkl<100> than for $1/2\hkl\,<111>$ screw dislocations, both dislocations appear relevant to rationalize plasticity in BCC Cr, as confirmed by experimental observations \cite{mclaren1964,reid1966,hale1969} which report activity for both $1/2\hkl[111]\,\hkl{110}$ and \hkl[100]\,\hkl{110} slip systems. We now focus on the competition between these two slip systems, comparing the lattice friction opposing glide of $1/2\,\hkl<111>$ and \hkl<100> screw dislocations.

From our \textit{ab initio} calculations, we find the structural properties of the two investigated types of screw dislocations to be weakly influenced by magnetism, except for the existence of a magnetic fault for the $1/2\,\hkl<111>$ screw dislocations in the AF phase as discussed in the previous section. We obtain lower energy barriers  opposing dislocation glide in \hkl{110} planes for \hkl<100> screw dislocations than for $1/2\,\hkl<111>$ in both phases. Also, the saddle point energy is lower in the AF phase for both $1/2\,\hkl<111>$ and \hkl<100> screw dislocations, indicating easier glide of dislocations in the AF than in the NM phase.

\begin{figure}[htb]
        \centering
        \includegraphics[width=\linewidth]{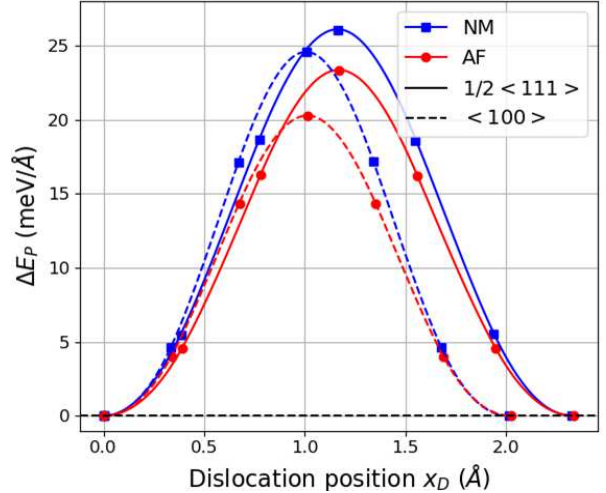}
        \caption{Peierls energy barriers for $1/2\,\hkl<111>$ and \hkl<100> screw dislocations gliding in a \hkl{110} plane in NM and AF phases as a function of the dislocation position $x_D=\xi \lambda_P$.}
        \label{fig:peierlsall}
\end{figure}

From the calculated Peierls barriers opposing dislocation glide in \hkl{110} planes, we can evaluate the Peierls stress $\tau_P$ for the investigated slip systems as
\begin{equation}
\tau_P=\dfrac{1}{b}\max\limits_{x_D} \dfrac{\partial E_P(x_D)}{\partial x_D}, \label{eq:peierls_stress}
\end{equation}
where $E_P$ is the Peierls potential, $x_D$ is the dislocation position in the glide plane, and $b$ is the norm of the Burgers vector. From the NEB calculations presented in Figures \ref{fig:peierls111} and \ref{fig:peierls100} for $1/2\,\hkl<111>$ and \hkl<100> screw dislocations respectively, the Peierls potential $E_P$ is known as a function of a reaction coordinate $\xi$ along the minimum energy path from one Peierls valley to the next nearest. However, the dislocation position is required in the above expression \ref{eq:peierls_stress} for the evaluation of the Peierls stress. As a first approximation, we assume that the dislocation position varies linearly with the reaction coordinate $\xi$ between two adjacent Peierls valleys separated by a distance $\lambda_P$ as $x_D=\xi \lambda_P$. More precise definitions of the dislocation positions from the stress variation are available \cite{dezerald2016,rodney2017,clouet2018,kraych2019}, but they require a different setup for the NEB calculations than the one used in sections \ref{sec:Peierls111} and \ref{sec:Peierls100}. Upon crossing of the barrier, $1/2\,\hkl<111>$ dislocations have to travel a higher distance than \hkl<100>, namely $a_0\sqrt{6}/3$ and $a_0\sqrt{2}/2$. We replot the Peierls barriers for the two considered systems in both phases in Figure \ref{fig:peierlsall} as a function of the approximated dislocation position.

The Peierls stresses $\tau_P$ obtained from Eq. \ref{eq:peierls_stress} for $1/2\,\hkl<111>$\,\hkl{110} and \hkl<100>\,\hkl{110} slip systems are 2.3 and 2.2\,GPa respectively in the NM phase, and 2.1 and 1.8\,GPa in the AF phase. We insist that these should be regarded as a rough estimation of the \textit{ab initio} Peierls stresses. Still, we find a close value for \hkl<100> and $1/2\,\hkl<111>$ screw dislocations in both phases, indicating overall an as easy glide for \hkl<100> and $1/2\,\hkl<111>$ dislocations. The difference between the two Peierls stresses is larger in the AF phase, in favor of an easier glide of \hkl<100> dislocations. But we have not considered here that $1/2\,\hkl<111>$ dislocations need to be paired in the AF phase, leading to \hkl<111> super-dislocations and probably lowering the associated Peierls stress.

\section{Conclusion}
This work investigates the impact of magnetism on the structural properties and mobility of screw dislocations in BCC Cr. We demonstrate the AF magnetic phase to be a good approximate of the SDW experimental ground state based on elastic and magnetic order considerations. The study of the generalized stacking fault energies along \hkl<111> directions revealed the introduction of a magnetic fault when the crystal is sheared in the AF phase by $1/2\,\hkl<111>$ in the three close-packed crystallographic planes \hkl{110}, \hkl{112} and \hkl{123} of the BCC lattice. As a consequence, $1/2\,\hkl<111>$ dislocations introduce a magnetic fault when shearing the crystal. Except for the introduction of this magnetic fault, our \textit{ab initio} modeling of the $1/2\,\hkl<111>$ screw dislocation evidences no structural difference between dislocation cores in the NM and AF phases, with also close core energies and Peierls energy barriers in the two magnetic phases. The main consequence of magnetism is the necessity for $1/2\,\hkl<111>$ dislocations to coexist and move pairwise, leading to super-dislocations with \hkl<111> Burgers vectors dissociated in two partial dislocations separated by a magnetic fault. 

\hkl<100> screw dislocations are found stable, with a compact core in both magnetic phases. The Peierls energy barrier opposing their glide in \hkl{110} planes has a slightly lower maximum than for $1/2\,\hkl<111>$ in both magnetic phases, leading to comparable Peierls stresses for the two dislocations. Our \textit{ab initio} study therefore demonstrates that both \hkl<111>\,\hkl{110} systems and \hkl<100>\,\hkl{110} have to be considered to describe Cr plasticity, in agreement with experiments showing activity for these two slip systems.

\section*{Acknowledgments}
This work was performed using HPC resources from GENCI-CINES and -TGCC (Grants 2019-096847)
and is funded by the French Tripartite Institute (CEA-EDF-Framatome) through the ICOMB project.
\bibliographystyle{elsarticle-num}
\biboptions{sort&compress}
\bibliography{ARTICLE}

\end{document}